\documentclass[aps,superscriptaddress,preprint,nofootinbib,eqsecnum]{revtex4}
\input{epsf}

\def\nc{N_{\rm c}}
\def\etacl{\eta_{\rm cl}}
\def\k{{\bf k}}
\def\p{{\bf p}}
\def\x{{\bf x}}
\def\pmax{p_{\rm max}}
\def\Eq#1{Eq.~(\ref{#1})}
\def\centerbox#1#2{\centerline{\epsfxsize=#1\textwidth\epsfbox{#2}}}
\usepackage{graphicx}

\begin{document}

\title{The stickiness of sound: An absolute lower limit on viscosity
and the breakdown of second order relativistic hydrodynamics}

\author{Pavel Kovtun}
\affiliation{
    Department of Physics and Astronomy,
    University of Victoria,
    Victoria, BC, V8P 5C2, Canada
}
\author{Guy D.\ Moore}
\affiliation{
    Department of Physics,
    McGill University,
    3600 rue University,
    Montr\'{e}al, QC H3A 2T8, Canada
}
\author{Paul Romatschke}
\affiliation{
    Frankfurt Institute for Advanced Studies,
    D-60438 Frankfurt, Germany 
}

\begin{abstract}
\noindent
Hydrodynamics predicts long-lived sound and shear waves.  Thermal
fluctuations in these waves can lead to the diffusion of momentum density,
contributing to the shear viscosity and other transport coefficients.
Within viscous hydrodynamics in 3+1 dimensions, this leads
to a positive contribution to the shear viscosity, which is finite but
inversely proportional to the microscopic shear viscosity.  Therefore
the effective infrared viscosity is bounded from below.  The
contribution to the second-order transport coefficient $\tau_\pi$ is
divergent, which means that second-order relativistic viscous
hydrodynamics is inconsistent below some frequency scale.  We estimate
the importance of each effect for the Quark-Gluon Plasma, finding them
to be minor if $\eta/s = 0.16$ but important if $\eta/s = 0.08$.
\end{abstract}

\date{Early April, 2011}

\maketitle

\section{Introduction}

Heavy ion collisions at RHIC \cite{Adcox:2004mh,Back:2004je,Arsene:2004fa,Adams:2005dq} and the LHC \cite{Aamodt:2010pa,Aamodt:2010pb}
produce a medium whose evolution, at least at early times, is well
described by hydrodynamics with a very small viscosity \cite{Kolb:2003dz,Hirano:2005xf,Huovinen:2006jp,LuzumRomatschke,Teaney:2009qa,Luzum:2009sb,Luzum:2010ag,Song:2011hk,Song:2011qa}
(at least when normalized to the entropy density \cite{KSS}).
A major theoretical goal is now to determine this viscosity as
accurately as possible, by modeling the development of a heavy ion
collision with viscous hydrodynamics.
In order to do so it is necessary to go beyond the
``first-order'' (Navier-Stokes) formalism, because the first-order
formalism leads to acausal and unstable evolution in a relativistic
setting \cite{Romatschke:2009im,Schafer:2009dj,Heinz:2009xj,Monnai:2010qp,Calzetta:2010au}.  As shown by Israel and
Stewart \cite{Muller,IsraelStewart,Hiscock}, this problem can be cured by
working instead with hydrodynamics expanded to the second order in
gradients, which should have the added advantage of
being more accurate.  Several groups have been involved in studying the
hydrodynamics of heavy ion collisions using such second-order formalisms
\cite{LuzumRomatschke,Dusling:2007gi,Song:2008si,Chaudhuri:2009hj,Schenke:2010nt,PeraltaRamos:2010je,Niemi:2011ix}.  However, within the community which studies
hydrodynamics and kinetics of atomic gases, it has been known for almost
40 years that the gradient expansion in hydrodynamics fails beyond the
first order \cite{SBE}.  Is this also true in the relativistic setting?
If so, what implications does it have for the study of hydrodynamics via
the second-order formalism?

\section{Setup and intuitive argument}

We start by reviewing relativistic hydrodynamics
to second order.  Hydrodynamics is the modeling of a fluid by solving
the stress-energy conservation equations%
\footnote{
    If there are conserved currents $J_a^\mu$ one also considers current
    conservation $\partial_\mu J_a^\mu = 0$.  In equilibrium
    $J_a^\mu = n_a u^\mu$, out of equilibrium there can be derivative
    corrections.  However, in ultra-relativistic heavy ion collisions
    the density of the conserved baryon number in the central rapidity
    region is small, so we will neglect it and will not discuss
    conserved currents further.
    }
\begin{equation}
\label{stressconserve}
\partial_\mu T^{\mu\nu}(x) = 0
\end{equation}
assuming some functional form for the stress tensor.  Ideal
hydrodynamics assumes the equilibrium form,
\begin{equation}
\label{Teq}
T_{\rm eq}^{\mu\nu}(x) = (\epsilon(x)+P)\, u^\mu(x)\, u^\nu(x)
   + P g^{\mu\nu} \,,
\qquad P = P(\epsilon)
\end{equation}
where $g_{\mu\nu} = {\rm Diag}[-1,+1,+1,+1]$ is the metric,
$u^\mu$ is the 4-velocity determining the rest-frame (normalized so
$u_\mu u^\mu = -1$), $\epsilon=u_\mu u_\nu T^{\mu\nu}$ is the rest-frame
energy density, and $P$ is the pressure as
determined by the equation of state $P=P(\epsilon)$.
Viscous hydrodynamics assumes that the fluid is near
equilibrium so that $T^{\mu\nu}$ is close to this form.  Assuming that
equilibration is a fast, local process, corrections to
this form can be written in terms of an expansion in gradients.  Israel
and Stewart showed \cite{IsraelStewart} that a slight re-organization of
the second-order derivative expansion yields stable equations which are
correct to second order provided a particular additional term appears at
second order;
\begin{equation}
\label{Tgradient}
T^{\mu\nu}(x) = T^{\mu\nu}_{\rm eq}+\Pi^{\mu\nu}\,,\qquad
\Pi^{\mu\nu}= - 2 \etacl \partial^{\langle \mu} u^{\nu \rangle}
 + \tau_\pi \left( u^\alpha \partial_\alpha
               \Pi^{\mu\nu}
       + \frac{1}{3} \Pi^{\mu\nu}
         \partial_\alpha u^\alpha \right)
\;\;\mbox{(+ other terms)}
\end{equation}
where $\etacl$ is the ``classical'' viscosity coefficient
that one would obtain from a microscopic calculation using
the Kubo formula based on (\ref{Tgradient}).
The angular brackets in $\partial^{\langle \mu} u^{\nu\rangle}$
mean that the indices are to be symmetrized, projected to be spatial in
the frame given by $u^\mu$, and trace-subtracted.%
\footnote{
    That is, defining $P^{\mu\nu} = g^{\mu\nu} + u^\mu u^\nu$ which
    is a projector to local rest-frame spatial components,
    $2\partial^{\langle\mu} u^{\nu\rangle} =
    \left(P^{\mu\alpha} P^{\nu\beta} + P^{\nu\alpha} P^{\mu\beta}
    - \frac{2}{3} P^{\mu\nu}
    P^{\alpha\beta}\right)\partial_\alpha u_\beta$.
    }
The extra terms
include bulk viscosity and nonlinear effects, and are catalogued in
\cite{BRSSS,Bhattacharyya:2008jc,REntropy}.  To simplify the
discussion here we will consider a conformal fluid, in which case the
bulk viscous term is absent.  While this is not a very good
approximation for QCD near the transition temperature, in practice the
inclusion of bulk viscosity and a realistic equation of state would have
only a small influence on our final results.%
\footnote{In particular our results are dominated by the effects of
  shear waves, which are not sensitive to bulk viscosity or the equation
  of state.}

Solving the hydrodynamic equations for small fluctuations in $u^\mu$,
$\epsilon$ about constant values, one finds two sorts of long-lived wave
solutions, sound waves and shear waves.  At lowest order in
$k,\omega \ll \epsilon / \etacl$, they obey dispersion relations of
\begin{equation}
\label{omegas}
\omega_{\rm shear} = -i \frac{\etacl}{\epsilon+P} k^2 \,, \qquad
\omega_{\rm sound} = \frac{\pm k}{\sqrt{3}}
                     - i \frac{2\etacl}{3(\epsilon+P)} k^2 \,.
\end{equation}
Each sort of wave decays with time, but with a decay rate which vanishes
quadratically in the small $k$ limit, as well as becoming small for
small $\etacl$.  Generically, out of equilibrium such waves will be
present with large amplitudes.  But equipartition of energy says that
even in equilibrium such waves will be present, carrying energy which
averages to $T/2$ per degree of freedom.  The long decay times of these
waves contradict the assumption that all degrees
of freedom in a fluid equilibrate via rapid local processes.
This imperils the assumption behind the gradient expansion in
\Eq{Tgradient}.  If there are arbitrarily slowly equilibrating degrees
of freedom, \Eq{Tgradient} can contain terms nonanalytic in gradients.
In the nonrelativistic setting it is known that precisely this happens
\cite{SBE}.

\begin{figure}
\centerbox{0.7}{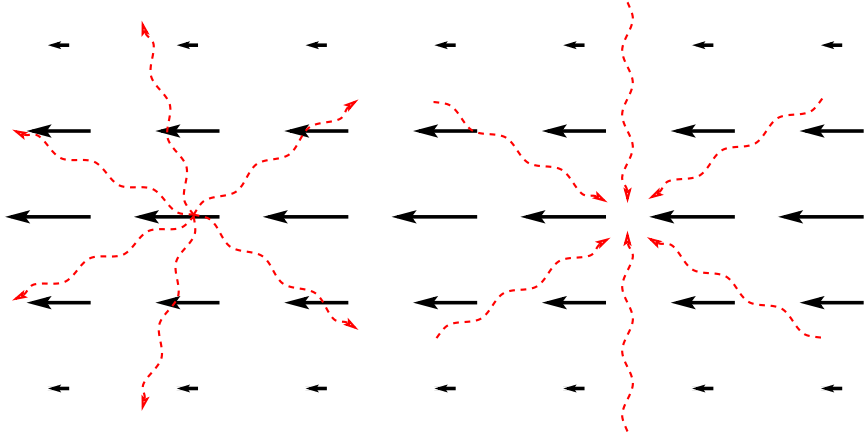}
\caption[Shear wave]
{\label{fig1}
 A shear wave; fluid moves to the left in a band of fluid near the
 middle of the figure.  Viscosity determines the loss (by diffusion) of
 the forward motion of this fluid.  Sound waves (dotted, in red) leaving
 the left-moving
 fluid carry, on average, net left-moving momentum, which is not
 compensated by sound waves arriving in the left-moving fluid.
 Hence sound waves contribute to viscosity.
}
\end{figure}

To see more intuitively how hydrodynamic waves can contribute to
hydrodynamic coefficients, consider a shear wave, as illustrated in
Figure \ref{fig1}.  Hydrodynamics says that this wave configuration will
decay with time as the $x$-momentum carried by the fluid diffuses in the
$y$ direction into neighboring regions, which are not flowing.  The rate of
this diffusion is controlled by shear viscosity, which is by definition
the diffusion coefficient for the component of momentum transverse to the
diffusion (the $y$-diffusion of $x$-momentum in this illustration).  But
one of the mechanisms which can transmit $x$-momentum is hydrodynamic
waves, such as sound waves, with wave lengths shorter than the
hydrodynamic structure considered.  Sound waves leaving the $x$-moving
fluid carry net $x$-momentum away, diffusing away this component of
momentum.  The phase space of such waves scales
as $d^3 k$ which is very UV dominated; but the distance propagated
before dissipation scales as $1/k^2$ as we just saw, leading to a
contribution to viscosity which scales as $\sim d^3 k / k^2$.  This is
IR finite, though it would not be in 2 spatial dimensions \cite{Forster:1977zz}.
The
contribution is larger for smaller $\etacl$, both because hydrodynamic
waves propagate further and because the range of $k$ where hydro is
valid expands at smaller $\etacl$, so the phase space of hydrodynamic
waves is larger.

Similarly, $\tau_\pi$ can be interpreted as a relaxation time; if a
fluid suddenly develops shear flow, $\tau_\pi$ is the time scale for
momentum diffusion to be established.  We already saw that hydrodynamic
modes contribute to momentum diffusion, with a mode of wave number $k$
contributing of order $1/k^2$.  The time scale for this mode to
leave equilibrium and establish its contribution to momentum diffusion
also scales as $1/k^2$.  This suggests that such a mode contributes to
$\tau_\pi$ by an amount proportional to $k^{-4}$, leading to an
$\int d^3 k / k^4$ contribution to $\tau_\pi$, which is small $k$
divergent.

\section{Computation of the contribution from Hydrodynamic waves}

Ref.~\cite{BRSSS} derives the following Kubo relation for the shear
viscosity $\etacl$ and the relaxation time $\tau_\pi$ in terms of the
retarded correlation function for two $T^{xy}$ stress-tensor operators:
\begin{equation}
\label{eq:Kubo}
  G_{\rm R}^{xy,xy}(\omega,k_z) = P - i\omega\etacl
                + \left(  \etacl\tau_\Pi - \frac{\kappa}{2} \right) \omega^2
                - \frac{\kappa}{2} k_z^2 + {\cal O}(\omega^3,k^3) \,.
\end{equation}
Here $\kappa$ is another transport coefficient discussed in
\cite{BRSSS}.  We see from the above that $\kappa$ can be extracted from
the zero frequency behavior of the stress-stress correlation function,
$G_{\rm R}^{xyxy}(\omega=0,k_z)$.  The zero frequency retarded function
equals the Euclidean correlation function, so $\kappa$ is a
thermodynamic property \cite{MooreSohrabi}, which is not sensitive to
long-wavelength hydrodynamic waves.  Since we are not interested in the
value of $\kappa$, we will consider $G_{\rm R}^{xy,xy}$ at
vanishing external spatial momentum $k=0$ and small nonzero frequency
$\omega$.

To compute the contribution of hydrodynamic waves to $\eta$ and
$\tau_\pi$, we compute their contribution to the above correlation
function, along the lines of what Kovtun and Yaffe did for the
symmetrized correlator \cite{KovtunYaffe}.  The operator $T^{xy}$ to be
used in the Green function above is the hydrodynamic one as described in
\Eq{Teq} and \Eq{Tgradient} above, allowing for thermally occupied
fluctuations in $u^\mu$ and $\epsilon$.  In the absence of
fluctuations $u^\mu = (1,0,0,0)$, $\epsilon=\epsilon_0$,
$P=\epsilon_0/3$, where $\epsilon_0$ is the equilibrium energy density.
Therefore $T^{xy}$ arises at order $(\delta u)$ due
to the viscous terms, and at order $(\delta u)^2$ and higher
from the equilibrium and viscous terms:
\begin{equation}
\label{Txy}
T^{xy} = \Big[ - \etacl (\partial^x u^y + \partial^y u^x ) \Big]
      + \Big[ (\epsilon{+}P) u^x u^y
              - \etacl (u^x \partial_0 u^y + u^y \partial_0 u^x)
              - \delta \epsilon \frac{d\etacl}{d\epsilon}
                (\partial^x u^y + \partial^y u^x) \Big]
      + {\cal O}(\delta^3) .
\end{equation}
The contribution at first order in fluctuations arises only from the
viscous term $-\etacl (\partial^x u^y + \partial^y u^x)$.
The {\em symmetrized} correlator is
\begin{equation}
G^{xyxy}_{\rm S}[\mbox{1-order}]
= \int d^3 x dt e^{i\omega t-i\k \cdot \x}
\left\langle\frac{1}{2}\left\{
  \Big( -\etacl [\partial^x u^y + \partial^y u^x](\x,t) \Big),
  \Big( -\etacl [\partial^x u^y + \partial^y u^x](0,0)  \Big)\right\}
\right\rangle \,,
\end{equation}
which is related to the retarded correlator via
$G_{\rm S}(\omega) = -[1+2n_b(\omega)] {\rm Im}\: G_{\rm R}(\omega)$.
The expression above is automatically ${\cal O}(k^2)$ and so it does not
contribute in the $k\rightarrow 0$ limit.  But there can also be local
contributions, that is, contributions proportional to $\delta^4(x)$ or
its derivatives, arising from the overlap of $T$ operators [contact
terms].  We will present a direct calculation of such contact terms
within hydrodynamic theory in a future publication \cite{to_appear}.
For our current purposes we will instead extract them by using stress
conservation,
\begin{equation}
\partial_\mu T^{\mu y} = 0 \quad \Rightarrow \quad
k_x^2 G^{xyxy}(k_x,\omega) = \omega^2 G^{0y0y} (k_x,\omega)
\end{equation}
together with the known expression for $G^{0y0y}_{\rm S}$
(see Eq.(33d) of \cite{KovtunYaffe}):
\begin{eqnarray}
G^{0y0y}_{\rm S}(k_x,\omega) &=& \frac{2k_x^2 \etacl T}
             {(k_x^2 \etacl/(\epsilon+P))^2 + \omega^2}
\qquad \mbox{and hence} \nonumber \\
G^{xyxy}_{\rm S}(k_x,\omega) & = & 2 \etacl T \left(
     1 - \frac{(k_x^2 \etacl/(\epsilon{+}P))^2}
              {(k_x^2 \etacl/(\epsilon{+}P))^2 + \omega^2} \right)\,.
\label{G_1order}
\end{eqnarray}
The first term in (\ref{G_1order}) is the contact term, which gives rise to the
$-i\omega \etacl$ term in \Eq{eq:Kubo}.  The second term is the
contribution from the correlator of the $-\etacl \partial^x u^y$ part of
$T^{xy}$, which vanishes at small $k$.

Now we extend this calculation to second order in fluctuations.
At this order there are higher order corrections to the terms involving
a single power of $\delta u$ in $T^{xy}$; and the lowest order
contribution from terms involving two powers of $u^x,u^y,\delta
\epsilon$ in $T^{xy}$.  But since the terms involving a single power of
$\delta u$ in $T^{xy}$ always involve spatial derivatives, these terms
all vanish in the $k \rightarrow 0$ limit, like the second term in
\Eq{G_1order} above.  So we skip their calculation and concentrate on
contributions from second-order in fluctuation terms in $T^{xy}$.
There are several such terms in \Eq{Txy} but we will concentrate on the
term $(\epsilon{+}P) u^x u^y$, and explain why the other terms can be
neglected at the end.  Since $u^x u^y$ is already quadratic in
fluctuations, we can neglect fluctuations in $\epsilon$ and replace
$(\epsilon{+}P)$ with $\epsilon_0{+}P_0 = \frac{4}{3} \epsilon_0$.
Since fluctuations in
$\epsilon$ will play no further role in the discussion we will
henceforth write $\epsilon,P$ for $\epsilon_0,P(\epsilon_0)$.

Define the correlation function of the fluid velocity $u^i$ to be
\begin{equation}
\label{Delta}
\Delta_{\rm S,R}^{ij}(\omega',\p)
\equiv \int dt\:d^3 \x \; e^{-i\p\cdot \x+i\omega t}
\langle u^i(t,\x) u^j(0,0) \rangle_{\rm S,R} \,,
\end{equation}
where $\langle\rangle_{\rm S,R}$ indicate whether the operators are to
be symmetrized (S) or if the retarded correlator is to be used (R).
Using the expression we have for the stress tensor above, the
symmetrized correlation function of two stress tensors at vanishing
external spatial momentum is
\begin{eqnarray}
\label{Gxyxy_S}
G^{xyxy}_{\rm S}[\mbox{2-order}](\omega,k=0) &=& (\epsilon+P)^2
\int \frac{d\omega'}{2\pi} \int \frac{d^3 \p}{(2\pi)^3}
\left( \Delta_{\rm S}^{xx}(\omega',\p)
       \Delta_{\rm S}^{yy}(\omega-\omega',-\p)
\vphantom{\Big|} \right. \nonumber \\
   && \hspace{1.52in} \left. \vphantom{\Big|}
     + \Delta_{\rm S}^{xy}(\omega',\p)
       \Delta_{\rm S}^{yx}(\omega-\omega',-\p) \right)\,, \qquad
\end{eqnarray}
where we assumed small, nearly linear hydrodynamic fluctuations and
small frequencies (details will be given in Ref.~ \cite{to_appear}).

We could use the KMS relation
$G_{\rm S}(\omega) = -[1+2n_b(\omega)] {\rm Im}\: G_{\rm R}(\omega)$
to extract the shear viscosity from this correlation function, but since
$\tau_\pi$ depends on the real part of the retarded function we would
need to invert the KMS condition through a Kramers-Kronig relation to get
the real part of $G_{\rm R}$.  It is more economical to compute
$G_{\rm R}$ directly;
\begin{eqnarray}
\label{Gxyxy_R}
G^{xyxy}_{\rm R}[\mbox{2-order}](\omega,k=0) &=& (\epsilon+P)^2
\int \frac{d\omega'}{2\pi} \int \frac{d^3 \p}{(2\pi)^3} \times
\nonumber \\ &&
\left( \Delta_{\rm S}^{xx}(\omega',\p)
       \Delta_{\rm R}^{yy}(\omega{-}\omega',-\p)
     + \Delta_{\rm S}^{xy}(\omega',\p)
       \Delta_{\rm R}^{yx}(\omega{-}\omega',-\p)
\vphantom{\Big|} \right. \nonumber \\ && \left. \vphantom{\Big|}
     + \Delta_{\rm R}^{xx}(\omega',\p)
       \Delta_{\rm S}^{yy}(\omega{-}\omega',-\p)
     + \Delta_{\rm R}^{xy}(\omega',\p)
       \Delta_{\rm S}^{yx}(\omega{-}\omega',-\p) \right),
\hspace{0.29in}
\end{eqnarray}
which in the small frequency limit reproduces $G_{\rm S}^{xyxy}$
when using the KMS condition.
This contribution is to be added to the first-order contribution, which,
as we discussed, reproduces the terms present in \Eq{eq:Kubo}.

Now $\Delta^{ij}_{\rm S}(\omega',\p)$ was determined by
Kovtun and Yaffe \cite{KovtunYaffe}.  They included the effect of the
viscous $\etacl$ term (first order gradients) but dropped the $\tau_\pi$ term
(second order gradients), a procedure we will
follow. One finds
\begin{eqnarray}
\label{DeltaS}
\Delta^{ij}_{\rm S}(\omega',\p)
 &=& \frac{2T}{\epsilon+P} \left[
  \frac{p^i p^j}{p^2} \frac{\tilde{\gamma}_\eta p^2 \omega^2}
                           {(\omega^2 - p^2/3)^2
                            + (\tilde{\gamma}_\eta p^2 \omega)^2}
  +\left( \delta^{ij} - \frac{p^i p^j}{p^2} \right)
    \frac{\gamma_\eta p^2}{(\gamma_\eta p^2)^2 + \omega^2}
  \right]
\\ \mbox{with} &&
\gamma_\eta \equiv \frac{\etacl}{\epsilon+P} \,, \quad
\tilde\gamma_\eta \equiv \frac{4 \etacl}{3(\epsilon+P)} \,.
\nonumber
\end{eqnarray}
The first and second terms in $\Delta^{ij}$ represent sound and shear
waves respectively.  This expression differs slightly from the one found
by Kovtun and Yaffe \cite{KovtunYaffe} because they assume
$p \gamma_\eta \ll 1$ which allows them to split the sound mode
contribution into separate terms for $\omega \simeq p/\sqrt{3}$
and $\omega \simeq -p/\sqrt{3}$ propagation.

In the small frequency limit, 
the factor $2\,T$ in the numerator of \Eq{DeltaS} should really be
interpreted as $\omega(1+2n_b(\omega))$, with
$n_b(\omega) = (e^{\omega/T}-1)^{-1}$ as usual.  Then the KMS relation
$\Delta^{ij}_S(\p,\omega)
   = -(1+2n_b(\omega)) {\rm Im}\: \Delta_{\rm R}^{ij}(\p,\omega)$
and the analytic properties of $\Delta_{\rm R}$ (no poles in the upper
half-plane) uniquely establish
\begin{equation}
\label{DeltaR}
\Delta^{ij}_{\rm R}(\p,\omega) = \frac{1}{\epsilon+P} \left[
  \frac{p^i p^j}{p^2} \frac{\omega^2}
                         {i \tilde\gamma_\eta p^2 \omega +(\omega^2-p^2/3)}
 +\left( \delta^{ij} - \frac{p^i p^j}{p^2} \right)
  \frac{-\gamma_\eta p^2}{-i \omega +\gamma_\eta p^2} \right] \,.
\end{equation}

We are now ready to compute \Eq{Gxyxy_R}. Since our calculation
is done entirely within hydrodynamics, or equivalently assuming
momenta to be small, our result will not be applicable at large momenta,
neither in the argument of $G^{xyxy}_{\rm R}$, nor for the hydrodynamic
propagators $\Delta^{ij}$ inside the integral of \Eq{Gxyxy_R}. Hence,
we need to restrict the calculation to the highest wave number $\pmax$,
or the inverse of the shortest length scale, where the
hydrodynamic description is valid.
Cutting off the $p$ integration at $\pmax$ and
considering first the shear-shear contribution we can perform the index contractions and angular integrals, finding
\begin{eqnarray}
\label{G_shearshear}
G^{xyxy}_{\rm R,shear{-}shear}(\omega) & = &
 \frac{14}{15} \frac{1}{2\pi^2} \int_0^{\pmax} p^2 dp
 \int \frac{d\omega'}{2\pi}
 \frac{2 \gamma_\eta p^2 T}
      {(\gamma_\eta p^2 -i \omega')(\gamma_\eta p^2 + i \omega')}
 \frac{-\gamma_\eta p^2}{\gamma_\eta p^2 - i \omega + i \omega'}
\nonumber \\
 & = & \frac{7 T}{30 \pi^2} \int_0^{\pmax} dp
     \frac{-p^4}{p^2 - i \omega / (2 \gamma_\eta)}
\nonumber \\
 & = & \frac{7 T}{30 \pi^2} \int_0^{\pmax} dp
   \left[ - p^2 - \frac{i \omega}{2 \gamma_\eta}
          + \frac{\omega^2 / (4 \gamma_\eta^2)}
                 {p^2 - i \omega / (2 \gamma_\eta)} \right]
\,.
\end{eqnarray}
The $p^2$ term is an uninteresting contribution of hydrodynamic waves to the
pressure.%
\footnote{%
    The pressure contribution has an unexpected sign.  In a weakly
    coupled theory, the contributions of ordinary particles in loops
    also have the wrong sign, which is over-canceled by a contact term,
    see \cite{RomatschkeSon}.}
The two ``interesting'' terms are
\begin{equation}
\label{midGxyxyresult}
G^{xyxy}_{\rm R,shear{-}shear}(\omega) \simeq
 -i \omega \frac{7 T \pmax}{60 \pi^2 \gamma_\eta}
 + (i+1) \omega^{\frac{3}{2}}
   \frac{7 T}{240 \pi \gamma_\eta^{\frac 32}} \,.
\end{equation}
Comparing with \Eq{eq:Kubo}, we see that the first term is
a positive contribution to the shear viscosity arising
from shear waves.  The
imaginary part of the second term is a frequency dependent reduction in
the shear viscosity, which vanishes as $\omega\rightarrow 0$.  Therefore
there is no problem defining the shear viscosity in terms of the zero
frequency limit of $\partial G^{xyxy}_{\rm R}/\partial \omega$.
The real part of the $\omega^{3/2}$ term has the same sign as
the $\tau_\pi$ term in \Eq{eq:Kubo}, but the wrong $\omega$
dependence. This term can be interpreted as a frequency dependent
correction to $\tau_\pi$ which diverges at small frequency; or it can be
interpreted as a breakdown of the validity of the hydrodynamic expansion
beyond one-derivative order.

We should also include the sound ($p^i p^j/p^2$) terms in $\Delta^{ij}$
shown in \Eq{DeltaS} and \Eq{DeltaR}.
If we assume $\pmax \gamma_\eta\ll 1$ then the sound part in (\ref{DeltaR}) may be
approximated as
$$
\frac{\omega^2}{i\tilde\gamma_\eta p^2 \omega +(\omega^2-p^2/3)}
\rightarrow \frac{\omega}{2}\left(
\frac{1}{\omega+i \tilde \gamma_\eta/2 p^2-p/\sqrt{3}}+\frac{1}{\omega+i \tilde \gamma_\eta/2 p^2+p/\sqrt{3}}\right)
$$
which has the advantage that now the $\omega^\prime$ poles in (\ref{Gxyxy_R})
are simple. The integration is then straightforward and for
the mixed shear-sound term one has
$$
\frac{-T}{5\pi^2}
\int_{-\pmax}^{\pmax} dp \frac{p^4}
            {p^2 - 3i (\omega-p/\sqrt{3})/5\gamma_\eta}
=-\frac{2T}{3\pi^2} \gamma_\eta^2 \pmax^5
+i \frac{2\gamma_\eta T}{3\pi^2} \omega \pmax^3 + \ldots \,.
$$
One thus finds that the mixed shear-sound term is suppressed with respect to the result
(\ref{midGxyxyresult}) by extra powers of $\pmax \gamma_\eta$.
The sound-sound term contains a part that has the same structure as the shear-shear contribution,
as well as other parts that are again suppressed by powers of $\pmax \gamma_\eta$.
Evaluation of the ``interesting'' contribution to leading order in $\pmax \gamma_\eta$
gives
\begin{equation}
\label{Mainresult}
G^{xyxy}_{\rm R}(\omega\ll \pmax \ll \gamma_\eta^{-1}) \simeq
 -i \omega \frac{17 T \pmax}{120 \pi^2 \gamma_\eta}
 + (i+1) \omega^{\frac{3}{2}}
   \frac{\left(7+\left(\frac{3}{2}\right)^{\frac 32} \right) T}
        {240 \pi \gamma_\eta^{\frac 32}} \,+{\cal O}(\pmax^2 \gamma_\eta^2,\omega^2)
\end{equation}
This is our main result.

Let us finally comment about whether or not we need to consider other
contributions to $G^{xyxy}_{\rm R}$ coming, for instance, from
the $-\etacl u^x \partial_0 u^y$ term in \Eq{Txy}.  Since hydrodynamics is
predicated on the convergence of the derivative expansion introduced in
Eqs.~(\ref{Teq},\ref{Tgradient}), $\pmax$ should be chosen as the
largest momentum scale where successive terms in the series are
successively smaller, which requires
$\pmax \ll (\epsilon+P)/\etacl$ (comparing the zero and one-derivative
terms), $\pmax \ll \etacl / (\etacl \tau_\pi) = \tau_\pi^{-1}$ (comparing
the one-derivative and two-derivative terms).  This also ensures that
hydrodynamic waves with $p < \pmax$ will have
${\rm Im}\:\omega \ll p$, so that hydrodynamic waves are well-defined,
long-lived excitations in the plasma.  And it ensures that the real
parts of the propagating frequencies of the two sound waves and the
shear wave are more widely separated than their imaginary parts.  These
conditions ensure that contributions to $G^{xyxy}_{\rm R}$ from
higher-derivative terms in \Eq{Tgradient}, and contributions arising
from interference between sound and shear waves in \Eq{Gxyxy_R}, are
small compared to the terms we have computed.  In particular, terms
arising from $T^{xy} \supset -\etacl u^x \partial_0 u^y$ in \Eq{Tgradient}
will  give rise to an integrand similar to \Eq{G_shearshear} but with
an extra power of $(p\gamma_\eta)^2$.  The resulting term will be
analytic in the frequency and will give corrections suppressed by
$\pmax^2 \gamma_\eta^2$ relative to the terms we have computed.
We will not attempt to compute such suppressed corrections here.

\section{Interpretation and Discussion}

\subsection{Viscosity}
\label{visc1}

We found that, besides the ``classical'' viscosity $\etacl$,
there is an additional contribution to viscosity
as measured on very long distance and time scales, generated by
relatively short-wavelength hydrodynamic waves. It is given 
by the coefficient of
$-i\omega$ in \Eq{Mainresult}:
\begin{equation}
\eta_{\rm new}=\frac{17 \pmax T(\epsilon{+}P)}{120 \pi^2 \etacl}\left(1 + {\cal O}(\pmax^2\gamma_\eta^2)\right)\,,
\end{equation}
where higher power corrections in $\pmax$ can be neglected as long as
$\pmax \gamma_\eta \ll 1$. 
The new contribution from \Eq{Mainresult} scales
as an inverse power of $\etacl$.  Therefore
$\eta \sim \etacl + {\cal O}(1/\etacl)$ 
has a positive minimum at a nonzero value of $\etacl$.
This places a lower
bound on the total (infrared) value of $\eta$ in any given theory.

To estimate the size of the new term and the minimum possible viscosity,
we need to estimate $\pmax$.  As discussed, $\pmax$ should the the largest
wave number such that the gradient expansion converges and a
hydrodynamic description is self-consistent.
Following the discussion after \Eq{Mainresult}, we will estimate
$\pmax \simeq \tau_\pi^{-1}/2$, or $\pmax \gamma_\eta \simeq \frac{1}{2}$.

The total
viscosity $\eta$ is then the classical plus the new contribution,
\begin{equation}
\eta = \etacl+\frac{17 \pmax \gamma_{\eta} T(\epsilon{+}P)^2}{120 \pi^2 \etacl^2}\,,
\end{equation}
which leads to the an absolute lower bound on the total viscosity:
\begin{equation}
\label{thebound}
\eta > \left(\frac{153}{160 \pi^2} T (\epsilon+P)^2 \pmax \gamma_\eta\right)^{1/3}\,.
\end{equation}

In a theory with
many colors and a weak coupling so that parametrically
\cite{PSS,AMY,YM}
\begin{equation}
\label{nc_est}
\epsilon \sim \nc^2 T^4 \,, \qquad
s \sim \nc^2 T^3 \,, \qquad
\eta \sim \nc^2 \alpha^{-2} T^3 \,, \qquad
\tau_\pi \sim \nc^0 \alpha^{-2} T^{-1} \,,
\end{equation}
the new contribution is parametrically
\begin{equation}
\eta_{\rm new} \sim \frac{\tau_\pi^{-1} \epsilon T}{\eta}
   \sim \alpha^4 T^3 \sim \alpha^{6} \nc^{-2} \eta
\end{equation}
showing that the new contribution is safely subdominant in any theory
which is either weakly coupled or has many fields.

However, in the real world QCD has 3 colors and the coupling at scales
of current interest is not small.  How large is this new contribution to
$\eta$ in this case?
To address this, we need estimates for $\etacl$,
$\tau_\pi$, and $\epsilon{+}P=sT$ for real-world QCD.
The easiest is $sT$, which can be determined on the lattice.
According to Bors\'anyi {\it et al} \cite{Fodor} Fig.~12, between
$T=200$MeV and $T=300$MeV $s/T^3$ rises from about 10 to about 14.
Bazavov {\it et al} \cite{Karsch} Fig.~8 gives comparable numbers, with
$s/T^3$ rising from 11 to 16 in the same range.  $\tau_\pi$ naturally
scales as $\etacl/sT=\gamma_\eta$, and estimates in QCD vary from
$2.6 \gamma_\eta$, the value in strongly coupled ${\cal N}{=}4$ SYM
\cite{BRSSS}, to$\sim 5\gamma_\eta$, the value in weakly coupled QCD
\cite{YM}.  The trend is that, while $\tau_\pi / \gamma_\eta$ is a pure
number of order unity, it is smaller in more strongly-coupled contexts
and larger in weakly-coupled contexts.  So in the context of QCD with a
small shear viscosity it probably makes sense to assume
$\tau_\pi/\gamma_\eta$ is at the low end of this estimated range.
Regarding $\etacl$, the lowest estimates for $\etacl/s$ are
around $0.08$, while $\etacl/s = 0.16$ may be on the high side in terms of
fitting elliptic flow data \cite{LuzumRomatschke}.

If we estimate that $\tau_\pi = 3\etacl/sT$, $\etacl/s=0.08$,
and $s=10T^3$, we find
$\pmax \sim 2T$ and $\eta_{\rm new} \sim 0.36 T^3 \sim .036 s$.
The true value of $\eta/s$ would then be $0.08 + 0.036 = 0.116$.
Varying the value of $\etacl/s$ while holding the other
estimates fixed, this is close to the minimum value of the {\em total}
viscosity.
On the other hand, if $\etacl = .16s$ but the other
estimates are the same, then $\pmax \sim 1T$ and
$\eta_{\rm new} \sim .09 T^3 \sim .01 s$, which is a negligible
correction to the total viscosity.  Therefore the importance of the
``new'' contribution to viscosity is quite sensitive to the value of
$\etacl/s$; for real-world values of other parameters 
$\eta/s=0.08$ appears to be impossible, but $\eta/s=0.16$ is not.

\subsection{Relaxation time $\tau_\pi$}

The presence of a nonanalytic term in the frequency expansion for
$G_{\rm R}^{xyxy}(\omega)$ implies that hydrodynamics at second order in
gradients does not, strictly speaking, work.  However, in practice we
are usually interested in applying hydrodynamics over some range of
time, with some limited time resolution, and with a finite accuracy
tolerance.  If the nonanalytic term is sufficiently small compared to
the $\omega^2$ term for the frequency scales which are of actual
importance in a particular problem, then there may be no issue in
practice with using second order hydrodynamics.  To see whether this is
the case, we should estimate the frequency $\omega_{\rm min}$ where the
$\omega^{\frac 32}$ term is larger than the $\etacl \tau_\pi \omega^2$
term. Clearly, for $\omega<\omega_{\rm min}$, the second-order treatment
becomes invalid.

Again using the parametric estimates of \Eq{nc_est},
the frequency expansion of $G^{xyxy}_{\rm R}$,
\Eq{eq:Kubo} plus \Eq{Mainresult}, is parametrically of form
\begin{eqnarray}
\label{parametric1}
G^{xyxy}_{\rm R}(\omega) & \sim & P - i (\etacl+\eta_{\rm new}) \omega
 + \etacl \tau_\pi \omega^2
\nonumber \\
& \sim & \nc^2 T^4 \omega^0
   -i \nc^2 \alpha^{-2} T^3 \omega^{1}
   +(1+i) \nc^{0} \alpha^{3} T^{\frac 52} \omega^{\frac 32}
   + \nc^2 \alpha^{-4} T^2 \omega^2 \,.
\end{eqnarray}
The frequency scale where the $\omega^{\frac 32}$ term first dominates
the $\omega^2$ term is parametrically
\begin{equation}
\nc^0 \alpha^3 T^{\frac 52} \omega^{\frac 32}
\sim \nc^2 \alpha^{-4} T^2 \omega^2
\quad \Longrightarrow \quad
\omega_{\rm min} \sim \nc^{-4} \alpha^{14} T \,.
\end{equation}
Therefore, in a theory which {\sl either} has a large number of colors
{\sl or} has weak coupling, the frequency scale where the new term
becomes important is parametrically tiny.  Then there is no obstacle to
using second-order hydrodynamics for $\omega>\omega_{\rm min}$;
the second-order treatment will only become
invalid at a frequency scale where it is in any case almost
irrelevant compared to the $\omega^{1}$ viscous term.

Now we turn to real-world QCD.  To determine the breakdown
scale of second-order hydrodynamics we again compute the scale where the
real part of the $\omega^{\frac 32}$ term in \Eq{Mainresult} equals the
$\etacl \tau_\pi \omega^2$ term in \Eq{eq:Kubo}:
\begin{equation}
\omega_{\rm min} = \frac{ (7 + (3/2)^{\frac 32})^2 T}{(240\pi)^2}
 \left( \frac{\etacl}{s} \right)^{-7}
 \left( \frac{\tau_\pi}{\gamma_\eta} \right)^{-2}
 \left( \frac{s}{T^3} \right)^{-2} \,.
\label{eq:breakdown}
\end{equation}

\begin{table}
\begin{tabular}{|c|c|c||c|} \hline
$\epsilon{+}P$ & $\tau_\pi$ & $\etacl/s$ & $\;\;\omega_{\rm min}$ $\;\;$ \\
\hline
$10 T^4$ & $3 \frac{\etacl}{sT}$ & 0.08 & $7 T$ \\
$10 T^4$ & $5 \frac{\etacl}{sT}$ & 0.08 & $2.6 T$ \\
$16 T^4$ & $3 \frac{\etacl}{sT}$ & 0.08 & $2.8 T$ \\
$10 T^4$ & $3 \frac{\etacl}{sT}$ & 0.16 & $T/18$ \\
$\;\; 16 T^4\;\;$ & $\;5 \frac{\eta}{sT}\;$ &
 $\;\;0.16\;\;$ & $T/125$ \\
\hline
\end{tabular}
\caption{\label{table:taupi}
 Frequency scale $\omega_{\rm min}$ where the $\omega^{\frac 32}$ term equals the
$\etacl \tau_\pi \omega^2$ term in $G^{xyxy}_{\rm R}(\omega)$.  For $\omega<\omega_{\rm min}$,
second-order hydrodynamics is inconsistent.}
\end{table}

We illustrate
the consequences for real-world QCD in Table \ref{table:taupi}, using
some of the estimates for $s/T^3$, $\tau_\pi$, and $\etacl$ discussed in
the last subsection to evaluate the frequency scale
$\omega_{\rm min}$.  Below this frequency second-order
hydrodynamics is certainly not consistent.  What we see is that, for
reasonable values of $s/T^3$ and of $\tau_\pi$, this frequency
scale is very low if $\etacl/s = 0.16$, but it is very high for
$\etacl/s = 0.08$.  Therefore, second-order hydrodynamics can be applied
to QCD if $\etacl/s \sim 0.16$ above the transition temperature; it breaks
down below frequency scales of order $T/20$, which corresponds to 
time scales above 20 Fermi, 
safely above the microphysical scale it is intended to
model in heavy-ion collisions (5-10 fm).  
But if $\etacl/s \sim 0.08$, there is a problem.  Second-order
hydrodynamics is then already inconsistent on frequency scales of order
$2.5T$, corresponding to time scales above 0.4 Fermi.  This is a shorter
time scale than our estimate $\pmax^{-1} \sim 1/2T$ for the shortest
scale on which hydrodynamics is to be reliable.  So in this case there
would be {\sl no} range of scales on which second-order hydrodynamics is
applicable.

\subsection{Viscosity --- again}

Let us accentuate the conclusions drawn in the previous subsection
by again considering the correction to the viscosity in (\ref{Mainresult}).
As discussed above, Eq.~(\ref{Mainresult}) contains a constant, $\pmax$
dependent correction to the viscosity, but its actual value depends on the
estimate for $\pmax$.

A different way to bound what values of viscosity are compatible with
second-order hydrodynamics can be obtained
by considering the imaginary part for
the $\omega^{3/2}$ term, which can be interpreted as a frequency-dependent
correction to the viscosity,
$$
\eta(\omega)=\etacl - \omega^{1/2} \frac{(7 + (3/2)^{\frac 32}) T}{(240 \pi)\gamma_\eta^{3/2}}\,.
$$
Note that the first and second term become of the same order at
$\omega=\omega_{\rm max}$,
where
$$
\omega_{\rm max}=\gamma_\eta^5 s^2\frac{(240 \pi)^2}{(7 + (3/2)^{\frac 32})^2}\,.
$$
Second-order hydrodynamics is certainly no longer applicable at
$\omega>\omega_{\rm max}$, because the frequency dependent contribution
to the viscosity becomes large. On the other hand, we found above
that second-order hydrodynamics also breaks down for
$\omega<\omega_{\rm   min}$, because there the frequency dependent
contribution to the relaxation time becomes dominant.

So for a second-order hydrodynamic description to be applicable at all,
there has to be a frequency window
$\omega_{\rm min}<\omega<\omega_{\rm max}$ which can only exist if
\begin{equation}
\label{viscbound2}
\etacl > \frac{ s^{\frac 23} T (7+(3/2)^{\frac 32})^{\frac 13}}
            { (240\pi)^{\frac 13} (\tau_\pi/\gamma_\eta)^{\frac 16}}
\end{equation}
This is a lower bound on viscosity if second-order hydrodynamics is to
have any range of validity.  If we require the range of
allowed frequencies $\omega$ to be non-zero, the viscosity
will have be even higher than this bound.

\begin{figure}
\includegraphics[width=.45\linewidth]{etabound.eps}
\hfill
\includegraphics[width=.45\linewidth]{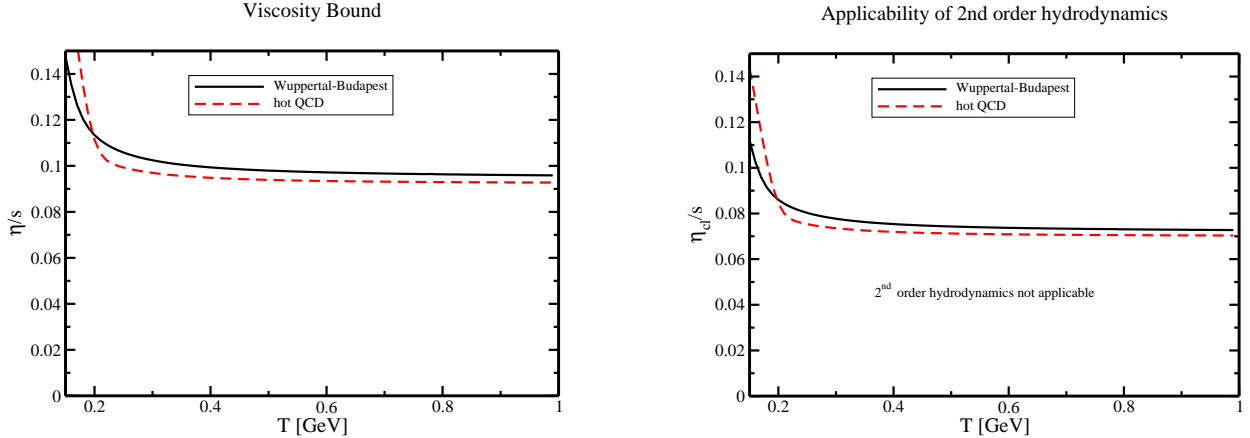}
\caption[Minvisc]
{\label{fig:minvisc}
Examples for the viscosity over entropy density bound (\ref{thebound})
(left) and applicability of second-order viscous hydrodynamics 
(\ref{viscbound2}) (right). The unknown parameters $\tau_\pi$
and $\pmax$ were assumed to be $\tau_\pi/\gamma_\eta=3$ and 
$\pmax=1/(2 \tau_\pi)$ and $s/T^3$ was evaluated using 
lattice QCD equations of state from the hotQCD \cite{Karsch} and Wuppertal-Budapest \cite{Fodor}
collaborations. 
}
\end{figure}

The temperature dependence of the minimal viscosity can be evaluated
when using lattice QCD data \cite{Fodor,Karsch} for the entropy $s$
and using $\tau_\pi/\gamma_\eta=3$, and is shown in Fig.~\ref{fig:minvisc}. 
We find that the minimal viscosity
as defined in \Eq{viscbound2} varies from about $\eta/s = 0.09$ at
$200$MeV to about $\eta/s=0.07$ in the Stefan-Boltzmann limit.  These
values are comparable to what we found is the lowest possible value of
$\eta/s$ in Subsection \ref{visc1}.

\section{Conclusions}

We have shown that, just as in non-relativistic hydrodynamics, in a
relativistic setting it is not self-consistent to consider the
hydrodynamic gradient expansion to second order.  The effects of thermal
fluctuations in the hydrodynamic variables themselves contribute to the
hydrodynamic evolution of the longest wavelength modes.  These effects
are suppressed both in weakly coupled theories and in theories with many
degrees of freedom such as QCD with many colors; but they can be
important in real-world QCD.

The correction
to the shear viscosity is positive and finite, so shear viscosity is a
well defined quantity in 3+1 dimensions.  However, in real-world QCD, if
$\etacl/s$
is very small then the hydrodynamic fluctuations can make a significant
additional contribution; estimating their size from our calculation,
it does not appear
to be possible for the shear viscosity to entropy ratio of real-world
QCD at $T=200$MeV to be smaller than $\eta/s \simeq 0.1$.

The issue becomes more severe for the second-order coefficient
$\tau_\pi$.  Strictly speaking this coefficient cannot be defined; its
definition assumes an analytic structure for the stress-stress
correlation function which is violated due to long-wavelength
hydrodynamic fluctuations.  In practice this is only really an issue at
long time scales (low frequencies), where the inclusion of second-order
effects may not be important anyway.  However, how long ``long'' is
depends on the value of the viscosity and the entropy density.
For real-world QCD, if $\etacl/s \sim 0.16$ then there is a wide range of
frequencies where the second-order theory is applicable; it
only fails at such low frequencies that the difference between
first-order and second-order hydro is insignificant at the affected
scales.  In this case there would be no real problem in practice
with using the second-order theory to {\sl e.g.\ } 
model heavy-ion collisions.  But if $\etacl/s \sim 0.08$, then
there is {\em no} range of scales where the application of second-order
hydrodynamics is consistent in real-world QCD.

While our calculation focussed on the case of relativistic 
hydrodynamics, a similar calculation should be possible in the
non-relativistic setting, presumably with qualitatively
similar results. We leave this interesting project for future work.

The effects we describe should not be present in dissipative
hydrodynamic simulations of heavy ion collisions as they are currently
conducted.  That is because, currently, such simulations include
dissipative viscous effects, but they do not include the fluctuations
required by the fluctuation-dissipation theorem to ensure that
hydrodynamic modes equilibrate with mean thermal excitation
amplitudes.  So the short-wavelength hydrodynamic waves responsible for
the effects we are discussing get quenched in existing hydrodynamic
simulations.  It would be very interesting to try to include
thermal fluctuations in hydrodynamic variables consistently in
hydrodynamic studies of heavy ion collisions.

\section*{Acknowledgements}

We would like to thank Derek Teaney
for useful conversations, and the organizers of the 2010 ESI workshop
'AdS Holography and the Quark-Gluon Plasma' in Vienna for providing
such a nice and stimulating environment that 
led to the initial conversations triggering this paper.
This work was supported in part
by the Natural Sciences and Engineering Research Council of Canada and
in part by the Helmholtz International Center for FAIR within the 
framework of the LOEWE program launched by the State of Hesse.


\begin{thebibliography}{99}


\bibitem{Adcox:2004mh}
  K.~Adcox {\it et al.}  [PHENIX Collaboration],
  {\it ``Formation of dense partonic matter in relativistic nucleus nucleus
  collisions at RHIC: Experimental evaluation by the PHENIX  collaboration,''}
  Nucl.\ Phys.\  A {\bf 757} (2005) 184
  [arXiv:nucl-ex/0410003].

\bibitem{Back:2004je}
  B.~B.~Back {\it et al.},
  {\it ``The PHOBOS perspective on discoveries at RHIC,''}
  Nucl.\ Phys.\  A {\bf 757} (2005) 28
  [arXiv:nucl-ex/0410022].

\bibitem{Arsene:2004fa}
  I.~Arsene {\it et al.}  [BRAHMS Collaboration],
  {\it ``Quark gluon plasma and color glass condensate at RHIC? The perspective
    from the BRAHMS experiment,''}
  Nucl.\ Phys.\  A {\bf 757} (2005) 1
  [arXiv:nucl-ex/0410020].

\bibitem{Adams:2005dq}
  J.~Adams {\it et al.}  [STAR Collaboration],
  {\it ``Experimental and theoretical challenges in the search for the quark  gluon
    plasma: The STAR collaboration's critical assessment of the  evidence from
  RHIC collisions,''}
  Nucl.\ Phys.\  A {\bf 757} (2005) 102
  [arXiv:nucl-ex/0501009].



\bibitem{Aamodt:2010pa}
  K.~Aamodt {\it et al.} [ The ALICE Collaboration ],
  {\it ``Elliptic flow of charged particles in Pb-Pb collisions at 2.76 TeV,''}

  [arXiv:1011.3914 [nucl-ex]].

\bibitem{Aamodt:2010pb}
  K.~Aamodt {\it et al.} [ The ALICE Collaboration ],
  {\it``Charged-particle multiplicity density at mid-rapidity in central Pb-Pb collisions at $\sqrt{s_{NN}}$ = 2.76 TeV,''}
  Phys.\ Rev.\ Lett.\  {\bf 105 } (2010)  252301.
  [arXiv:1011.3916 [nucl-ex]].




\bibitem{Kolb:2003dz}
  P.~F.~Kolb, U.~W.~Heinz,
  {\it ``Hydrodynamic description of ultrarelativistic heavy ion collisions,''}
  In *Hwa, R.C. (ed.) et al.: Quark gluon plasma* 634-714.
  [nucl-th/0305084].

\bibitem{Hirano:2005xf}
  T.~Hirano, U.~W.~Heinz, D.~Kharzeev, R.~Lacey, Y.~Nara,
  {\it ``Hadronic dissipative effects on elliptic flow in ultrarelativistic heavy-ion collisions,''}
  Phys.\ Lett.\  {\bf B636 } (2006)  299-304.
  [nucl-th/0511046].


\bibitem{Huovinen:2006jp}
  P.~Huovinen, P.~V.~Ruuskanen,
  {\it ``Hydrodynamic Models for Heavy Ion Collisions,''}
  Ann.\ Rev.\ Nucl.\ Part.\ Sci.\  {\bf 56 } (2006)  163-206.
  [nucl-th/0605008].

\bibitem{LuzumRomatschke}
M.~Luzum and P.~Romatschke,
  {\it ``Conformal Relativistic Viscous Hydrodynamics: Applications to RHIC results
  at $\sqrt{s_{NN}}$ = 200 GeV,''}
  Phys.\ Rev.\  C {\bf 78}, 034915 (2008)
  [Erratum-ibid.\  C {\bf 79}, 039903 (2009)]
  [arXiv:0804.4015 [nucl-th]].

\bibitem{Teaney:2009qa}
  D.~A.~Teaney,
  {\it ``Viscous Hydrodynamics and the Quark Gluon Plasma,''}
  %
  [arXiv:0905.2433 [nucl-th]].


\bibitem{Luzum:2009sb}
  M.~Luzum, P.~Romatschke,
  {\it ``Viscous Hydrodynamic Predictions for Nuclear Collisions at the LHC,''}
  Phys.\ Rev.\ Lett.\  {\bf 103 } (2009)  262302.
  [arXiv:0901.4588 [nucl-th]].

\bibitem{Luzum:2010ag}
  M.~Luzum,
  {\it ``Elliptic flow at LHC: Comparing heavy ion data to viscous hydrodynamic prediction,'}
  %
  [arXiv:1011.5173 [nucl-th]].



\bibitem{Song:2011hk}
  H.~Song, S.~A.~Bass, U.~W.~Heinz, T.~Hirano, C.~Shen,
  {\it ``Hadron spectra and elliptic flow for 200 A GeV Au+Au collisions from viscous hydrodynamics coupled to a Boltzmann cascade,''}
  %
  [arXiv:1101.4638 [nucl-th]].

\bibitem{Song:2011qa}
  H.~Song, S.~A.~Bass, U.~W.~Heinz,
  {\it ``Elliptic flow in 200 A GeV Au+Au collisions and 2.76 A TeV Pb+Pb collisions: insights from viscous hydrodynamics + hadron cascade hybrid model,''}
  %
  [arXiv:1103.2380 [nucl-th]].



\bibitem{KSS}
  P.~Kovtun, D.~T.~Son and A.~O.~Starinets,
  {\it ``Viscosity in strongly interacting quantum field theories from black hole physics,''}
  Phys.\ Rev.\ Lett.\  {\bf 94}, 111601 (2005)
  [arXiv:hep-th/0405231].





\bibitem{Romatschke:2009im}
  P.~Romatschke,
  {\it ``New Developments in Relativistic Viscous Hydrodynamics,''}
  Int.\ J.\ Mod.\ Phys.\  {\bf E19 } (2010)  1-53.
  [arXiv:0902.3663 [hep-ph]].

\bibitem{Schafer:2009dj}
  T.~Schafer, D.~Teaney,
  {\it ``Nearly Perfect Fluidity: From Cold Atomic Gases to Hot Quark Gluon Plasmas,''}
  Rept.\ Prog.\ Phys.\  {\bf 72 } (2009)  126001.
  [arXiv:0904.3107 [hep-ph]].

\bibitem{Heinz:2009xj}
  U.~W.~Heinz,
  {\it ``Early collective expansion: Relativistic hydrodynamics and the transport properties of QCD matter,''}
   [arXiv:0901.4355 [nucl-th]].

\bibitem{Monnai:2010qp}
  A.~Monnai, T.~Hirano,
  {\it ``Relativistic Dissipative Hydrodynamic Equations at the Second Order for Multi-Component Systems with Multiple Conserved Currents,''}
  Nucl.\ Phys.\  {\bf A847 } (2010)  283-314.
  [arXiv:1003.3087 [nucl-th]].

\bibitem{Calzetta:2010au}
  E.~Calzetta, J.~Peralta-Ramos,
  {\it ``Linking the hydrodynamic and kinetic description of a dissipative relativistic conformal theory,''}
  Phys.\ Rev.\  {\bf D82 } (2010)  106003.
  [arXiv:1009.2400 [hep-ph]].


\bibitem{Muller}
  I.\ M\"uller, Z.\ Phys.\ {\bf 198}, 329 (1967).


\bibitem{IsraelStewart}
W.~Israel,
  {\it ``Nonstationary Irreversible Thermodynamics: A Causal Relativistic Theory,''}
  Annals Phys.\  {\bf 100}, 310 (1976);
W.~Israel and J.~M.~Stewart,
  {\it ``Transient relativistic thermodynamics and kinetic theory,''}
  Annals Phys.\  {\bf 118}, 341 (1979).

\bibitem{Hiscock}
W.~A.~Hiscock and L.~Lindblom,
  {\it ``Stability and causality in dissipative relativistic fluids,''}
  Annals Phys.\  {\bf 151}, 466 (1983);
  Phys.\ Rev.\ {\bf D31} 725 (1985);
  Phys.\ Rev.\ {\bf D35} 3723 (1987);
  Phys. Lett. {\bf A 131} 509 (1988).

\bibitem{Dusling:2007gi}
  K.~Dusling, D.~Teaney,
  {\it ``Simulating elliptic flow with viscous hydrodynamics,''}
  Phys.\ Rev.\  {\bf C77 } (2008)  034905.
  [arXiv:0710.5932 [nucl-th]].

\bibitem{Song:2008si}
  H.~Song, U.~W.~Heinz,
  {\it ``Multiplicity scaling in ideal and viscous hydrodynamics,''}
  Phys.\ Rev.\  {\bf C78 } (2008)  024902.
  [arXiv:0805.1756 [nucl-th]].


\bibitem{PeraltaRamos:2010je}
  J.~Peralta-Ramos, E.~Calzetta,
  {\it ``Divergence-type 2+1 dissipative hydrodynamics applied to heavy-ion collisions,''}
  Phys.\ Rev.\  {\bf C82 } (2010)  054905.
  [arXiv:1003.1091 [hep-ph]].

\bibitem{Niemi:2011ix}
  H.~Niemi, G.~S.~Denicol, P.~Huovinen, E.~Molnar, D.~H.~Rischke,
  {\it ``Influence of the shear viscosity of the quark-gluon plasma on elliptic flow in ultrarelativistic heavy-ion collisions,''}

  [arXiv:1101.2442 [nucl-th]].

\bibitem{Schenke:2010nt}
  B.~Schenke, S.~Jeon, C.~Gale,
  {\it ``(3+1)D hydrodynamic simulation of relativistic heavy-ion collisions,''}
  Phys.\ Rev.\  {\bf C82 } (2010)  014903.
  [arXiv:1004.1408 [hep-ph]].

\bibitem{Chaudhuri:2009hj}
  A.~K.~Chaudhuri,
  {\it ``Centrality dependence of elliptic flow and QGP viscosity,''}
  J.\ Phys.\ G {\bf G37 } (2010)  075011.
  [arXiv:0910.0979 [nucl-th]].



\bibitem{SBE}
  I.~M.~De~Schepper, H.~Van~Beyeren and M.~H.~Ernst,
  {\it ``The nonexistence of the linear diffusion equation beyond Fick's law,''}
  Physica {\bf 75}, 1 (1974).

\bibitem{BRSSS}
  R.~Baier, P.~Romatschke, D.~T.~Son, A.~O.~Starinets and M.~A.~Stephanov,
  {\it ``Relativistic viscous hydrodynamics, conformal invariance, and holography,''}
  JHEP {\bf 0804}, 100 (2008)
  [arXiv:0712.2451 [hep-th]].

\bibitem{Bhattacharyya:2008jc}
  S.~Bhattacharyya, V.~E.~Hubeny, S.~Minwalla, M.~Rangamani,
  {\it ``Nonlinear Fluid Dynamics from Gravity,''}
  JHEP {\bf 0802 } (2008)  045.
  [arXiv:0712.2456 [hep-th]].


\bibitem{REntropy}
 P.~Romatschke,
  {\it ``Relativistic Viscous Fluid Dynamics and Non-Equilibrium Entropy,''}
  Class.\ Quant.\ Grav.\  {\bf 27}, 025006 (2010).
  [arXiv:0906.4787 [hep-th]].


\bibitem{Forster:1977zz}
  D.~Forster, D.~R.~Nelson, M.~J.~Stephen,
  {\it ``Large-distance and long-time properties of a randomly stirred fluid,''}
  Phys.\ Rev.\  {\bf A16 } (1977)  732-749.


\bibitem{MooreSohrabi}
G.~D.~Moore and K.~A.~Sohrabi,
  {\it ``Kubo Formulae for Second-Order Hydrodynamic Coefficients,''}
  arXiv:1007.5333 [hep-ph].

\bibitem{KovtunYaffe}
P.~Kovtun and L.~G.~Yaffe,
  {\it ``Hydrodynamic fluctuations, long-time tails, and supersymmetry,''}
  Phys.\ Rev.\  D {\bf 68}, 025007 (2003)
  [arXiv:hep-th/0303010].

\bibitem{to_appear} P.\ Kovtun, G.\ D.\ Moore, and P.\ Romatschke,
  to appear.

\bibitem{RomatschkeSon}
P.~Romatschke and D.~T.~Son,
  {\it ``Spectral sum rules for the quark-gluon plasma,''}
  Phys.\ Rev.\  D {\bf 80}, 065021 (2009)
  [arXiv:0903.3946 [hep-ph]].

\bibitem{PSS}
  G.~Policastro, D.~T.~Son and A.~O.~Starinets,
  {\it ``The shear viscosity of strongly coupled N = 4 supersymmetric Yang-Mills plasma,''}
  Phys.\ Rev.\ Lett.\  {\bf 87}, 081601 (2001)
  [arXiv:hep-th/0104066].

\bibitem{AMY}
  P.~B.~Arnold, G.~D.~Moore and L.~G.~Yaffe,
  {\it ``Transport coefficients in high temperature gauge theories: (I) Leading-log results,''}
  JHEP {\bf 0011}, 001 (2000)
  [arXiv:hep-ph/0010177].


\bibitem{YM}
  M.~A.~York and G.~D.~Moore,
  {\it ``Second order hydrodynamic coefficients from kinetic theory,''}
  Phys.\ Rev.\  D {\bf 79}, 054011 (2009)
  [arXiv:0811.0729 [hep-ph]].

\bibitem{Fodor}
S.~Bors\'anyi {\it et al.},
  {\it ``The QCD equation of state with dynamical quarks,''}
  JHEP {\bf 1011}, 077 (2010)
  [arXiv:1007.2580 [hep-lat]].

\bibitem{Karsch}
A.~Bazavov {\it et al.},
  {\it ``Equation of state and QCD transition at finite temperature,''}
  Phys.\ Rev.\  D {\bf 80}, 014504 (2009)
  [arXiv:0903.4379 [hep-lat]].

\end{thebibliography}
\end{document}